# Simulating the Impact of a Molecular 'Decision-Process' on Cellular Phenotype and Multicellular Patterns in Brain Tumors

**Chaitanya Athale [1], Yuri Mansury [1] and Thomas S. Deisboeck [1,*]**

[1] Complex Biosystems Modeling Laboratory, Harvard-MIT (HST) Athinoula A. Martinos Center for Biomedical Imaging, Massachusetts General Hospital, Charlestown, MA 02129.

**\*Corresponding Author:**

Thomas S. Deisboeck, M.D.
Complex Biosystems Modeling Laboratory
Harvard-MIT (HST) Athinoula A. Martinos Center for Biomedical Imaging
Massachusetts General Hospital-East, 2301
Bldg. 149, 13th Street
Charlestown, MA 02129
Tel: 617-724-1845
Fax: 617-726-5079
Email: deisboec@helix.mgh.harvard.edu

# Abstract

Experimental evidence indicates that human brain cancer cells proliferate *or* migrate, yet do not display both phenotypes at the same time. Here, we present a novel computational model simulating this cellular *decision-process* leading up to either phenotype based on a molecular interaction network of genes and proteins. The model's regulatory network consists of the epidermal growth factor receptor (EGFR), its ligand transforming growth factor-α (TGFα), the downstream enzyme phospholipaseC-γ (PLCγ) and a mitosis-associated response pathway. This network is activated by autocrine TGFα secretion, and the EGFR-dependent downstream signaling this step triggers, as well as modulated by an extrinsic nutritive glucose gradient. Employing a framework of mass action kinetics within a multiscale agent-based environment, we analyze *both* the emergent multicellular behavior of tumor growth and the single-cell molecular profiles that change over time and space. Our results show that one can indeed simulate the *dichotomy* between cell migration and proliferation based *solely* on an EGFR decision network. It turns out that these behavioral decisions on the single cell level impact the spatial dynamics of the entire cancerous system. Furthermore, the simulation results yield intriguing experimentally testable hypotheses also on the sub-cellular level such as spatial cytosolic *polarization* of PLCγ towards an extrinsic chemotactic gradient. Implications of these results for future works, both on the modeling and experimental side are discussed.

# 1. Introduction

This paper proposes a model of *gene-protein interactions* integrated in an agent-based system. We use the system to simulate the ability of cancer cells to 'switch' between migrating and proliferating phenotypes and argue that this molecular 'decision-process' is capable of reproducing some of the experimentally observed, multicellular spatio-temporal dynamics of brain tumor expansion. The smallest unit of our model is a molecular species which interacts with other molecules within and across sub-cellular compartments as well as with local microenvironment. The dynamic change in concentration of these molecular species both inside and around the tumor cell guides its phenotypic behavior. Therefore, a particular novel feature of our study here is the explicit modeling of the *feedback* effects from molecular-level dynamics into cellular behavior. This single-cell decision in turn affects the overall tumor growth dynamics and as such yields a truly *multi-scale cancer model*.

## 1.1. Dichotomy of Glioma Cells

For highly malignant brain tumors, i.e., gliomas, Giese et al. (1996) first proposed *dichotomy* between the *phenotypes* of migration and proliferation in such cells. The authors argue that these two fates appear to exclude each other such that cells that proliferate do not migrate and vice-a-versa. However, the exact molecular mechanism governing this suggested *switch* has not yet been clearly established. Tissue cell invasiveness of gliomas is considered a major reason for the poor outcome of patients suffering from the disease (reviewed in Berens and Giese, 1999), emphasizing the need to better understand the tumor biology governing such dichotomy. Our model aims to address some of these issues by examining the spatio-temporal dynamics of the molecular processes that guide the decision between motile and proliferative *traits* of a cancerous cell, and hence determine multicellular tumor growth dynamics. In our model, we choose to investigate the role of the *epidermal growth factor (EGF) receptor (EGFR)-mediated signaling pathway* since both, i*n vitro* and *in vivo* experiments with glioma cells have shown it to be involved in both the proliferative as well as the migratory response (Chicoine and Silbergeld, 1997).

## 1.2.    Previous Works on EGFR-Pathway Modeling

The effect of EGFR activation on cancer cells is diverse and complex (Prenzel et al., 2000) yet earlier studies of *molecular interaction* models in *single* cells have already examined various quantitative aspects of this multi-functional pathway. For instance, a model on the mitotic effect of ligand-based EGFR stimulation has correlated DNA synthesis with receptor occupancy (Wiley and Cunningham, 1981). Lauffenburger and Linderman (1996) then incorporated the linear relationship of cell proliferation in response to EGFR occupancy into a phenomenological model, which we use in our work here. Endocytosis is a major regulator of EGFR receptor trafficking (Resat et al., 2003; Starbuck and Lauffenburger, 1992) and influences signaling. Moreover, Brightman and Fell (2000) demonstrated that the differences in effect of EGF and nerve growth factor (NGF) on the same network were due to differential feedback. Such diverse effects of EGFR signaling are due mainly to the dynamics of downstream pathways as shown by Shvartsman et al. (2002a). Interestingly, a recent and detailed model has demonstrated the robustness of this network to large variations in initial values (Schoeberl et al., 2002). Furthermore, spatial 2D models of EGFR signaling have also examined pattern generation with *multiple* cells (Shvartsman et al., 2002b) as well as self-organization of spatial-polarization in single cells in 2D that use EGFR-like autocrine signaling (Maly et al., 2004). Additionally, a study of the spatial range of *autocrine* signaling in the EGFR system established the rapid and local nature of autocrine ligand capture (Shvartsman et al., 2001). It thus noteworthy that paracrine and juxtacrine signaling by TGFα (as the EGFR ligand) has been modeled in the past using linearized coupled ordinary differential equations (Owen and Sherratt, 1998) and shown to be applicable to pattern formation (Owen et al., 2000). We incorporate several components from these previous concepts in our model, as will be described in the subsequent sections.

### 1.3.    Previous Works on Tumor Modeling

Briefly, earlier computational models of tumor growth have focused exclusively on *either* migratory *or* proliferative behavior. On the migratory side, models for instance examined oscillations in the invasive speed (Perumpanani et al., 1996) and obtained invasiveness parameters by model-based analysis (Tracqui et al., 1995). Other works focused on proliferative behavior. For example, a reaction-diffusion and pH-based model was used to examine the transition from benign to malignant cancer (Gatenby and Gawlinski, 1996). Another model studying spheroid growth involving positive feedback initiated by cell-cell interactions showed improved fits to experimental data (Marusic et al., 1991). Important for our efforts here, more recent approaches have *combined* both invasion *and* proliferation. These include efforts to fit the model to patient data on tumor growth dynamics (Tracqui, 1995), predict three dimensional dynamics using microscopic parameters (Kansal et al., 2000), employ differential diffusion in regions of the brain (Swanson et al., 2000) and incorporate cellular physiology like mitosis, apoptosis, necrosis, and nutrient uptake (Dormann and Deutsch, 2002). Our own agent-based *'microscopic-macroscopic'* brain tumor model, which includes proliferation and migration as well as cellular physiology (Mansury et al., 2002) has recently been extended to include a simplified network of two interacting genes, TenascinC and proliferating cell nuclear antigen (PCNA) that correlate with the cell phenotype of migration and proliferation, respectively (Mansury and Deisboeck, 2004b). Using this *multicellular* framework, in here we simulate the phenotypic switch from cell migration to proliferation and vice-a-versa, and examine its effects on growth factor- and nutrient dependent tumor growth dynamics while maintaining a *molecular* "resolution" on the single cell level.

## 1.4.    EGFR-Pathway Modeling Concept and Experimental Evidence

We model the EGFR-mediated signaling pathway as a network of interacting genes and proteins[1]. This pathway has been demonstrated to cause *invasiveness* in three human glioma cell lines in co-culture with fetal rat brain aggregates (Penar et al., 1997), a result confirmed by activation of EGFR in cultured glioma cells by autocrine transforming growth factor-α (TGFα), which led to both increased migration and scattering (El-Obeid et al., 1997). On the other hand, *proliferation* responses in gliomas have been observed in culture studies of EGFR activation with TGFα *in vitro* (Rubenstein et al., 2001), within a co-culture system (El-Obeid et al., 2002), in tissue samples from patients (von Bossanyi et al., 1998) and with external EGF-stimulation in cell culture (Li et al., 2003). Taken together, these findings suggest that EGFR activity itself is ambiguous for deciding the phenotype of the cell. We then argue that *differential processing* of the signal downstream in the EGFR-cascade (Wells, 1999) can *cause* the *phenotypic decision*.

*Role of EGFR Downstream Pathways:* The EGFR signaling pathway has been implicated in numerous downstream pathways, both in fibroblasts (Wells, 1999) as well as gliomas (Besson and Yong, 2001; Mischel and Cloughesy, 2003). It mainly affects the following two cascades, i.e. (*i*) PLCγ-Protein-Kinase C, and (*ii*) ERK-MAPK, which in turn affect multitudes of other pathways downstream (Besson and Yong, 2001). PLCγ activation dynamics have already been implicated in the switch of cellular behavior as suggested by literature on EGFR signaling (Chen et al., 1994; Piccolo et al., 2002; Wang et al., 1998; Wells et al., 1999).

Specifically, EGFR mediated PLCγ activity is necessary for cell motility in experiments with U87 glioma spheroids (Khoshyomn et al., 1999) as well as breast cancer cells *in vitro* (Kruger and Reddy, 2003). Interestingly, mitosis due to EGFR activation was inhibited by PLCγ in fibroblasts (Chen et al., 1994; Chen et al., 1996). Finally, recent evidence has shown that subtle differences in the dynamics of PLCγ activation appear to cause the different behavioral responses of the tumor cell to EGFR signaling. Specifically, a transient increase in PLCγ causes migration whereas a sustained, lower-level activation causes proliferation in human breast cancer cells (Dittmar et al., 2002). Thus signal discrimination is likely to be due to phospholipaseC-γ (PLCγ) which is activated downstream of EGFR and shows feedback inhibition of EGFR activity (reviewed in Wells (1999)). We therefore focus in here on the PLCγ-pathway and detail our concept in the following section.

### 1.4.1. Cellular 'Decision Making'

Experimental evidence suggests the following scenario for cellular behavior:

- Glioma cells continually produce basal levels of TGFα while their EGFR pathway is active. This has been demonstrated with immuno-staining methods in 88% of malignant gliomas of which a random cell subset is actively *proliferating* (Maruno et al., 1991; van der Valk et al., 1997). The proliferative cellular state can be changed by an external trigger, e.g. in here, through a diffusing glucose concentration in the cell's microenvironment, followed by rapid uptake of the nutrient and increased phosphorylation of the TGFα-EGFR activated complex (Hertel et al., 1986; Steinbach et al., 2004).

- This step now leads to a rapid increase in the active TGFα-EGFR complex resulting in a transient peak in PLCγ, which then triggers the activation of cell *migration* (Dittmar et al., 2002). Based on a direction-sensing mechanism that is guided by the peak concentration of active PLCγ, the glioma cell performs *chemotaxis* towards a nutritive site (von Bulow et al., 2001), again, represented in our case by the aforementioned replenished source of glucose.

- However, increasing PLCγ activation also reduces TGFα-EGFR activation through *negative feedback* (Chen et al., 1994; Chen et al., 1996; Wells, 1999) which then generates a *proliferative signal* (Dittmar et al., 2002; Kruger and Reddy, 2003). The extent of this proliferation signal is linearly dependent on the TGFα concentration (Maruno et al., 1991).

In summary, the (virtual) tumor cell in our model therefore *migrates* if the PLCγ molecule is *transiently* induced by TGF-α-dependent activation of EGFR and modulation by glucose, and *proliferates* if PLCγ is activated in a *sustained* manner by EGFR activation. This latter state is assumed to activate the ERK-MAPK pathway. Since ERK-MAPK signaling has been shown to be implicated in the proliferative response (Chajry et al., 1994) we treat this pathway implicitly, by replacing it with a signal for cell proliferation.

To model the entire network, we extend our previously developed agent-based modeling framework (Mansury and Deisboeck, 2003; Mansury and Deisboeck, 2004a, 2004b; Mansury et al., 2002) and add a novel intracellular network module. In the following section, we describe the modeling algorithm in detail.

# 2. Mathematical Model

Each autonomous agent or cancer cell is itself made up of three 'sub-cellular' *compartments*, i.e., nucleus, cytoplasm and membrane (**Figure 1a**). These compartments are further resolved into spatial *sub*-compartments directed towards the cardinal directions, i.e., North, South, East and West of the grid, and are each connected to two other neighboring compartments by rates of *in-* and *out-flows* (**Figure 1b**).

**Figure 1 a**

**Figure 1 b**

Each sub-compartment contains all the molecules involved in the EGFR signaling network (**Figure 2**). Mass balance reactions govern the flux of molecules from one sub-compartment to another, as well as reactions defined by the interaction network. Local autocrine secretion of TGFα, a concentration profile of glucose and the spatial restriction due to other cells in neighboring grid positions form the microenvironment of each cell (see also section **2.3.**). Thus, there is both chemoattraction by glucose *and* by TGFα. The latter has been shown in mammalian systems to be captured rapidly by cells, and therefore can act in an autocrine and juxtacrine (affecting neighboring cells) manner (Kempiak et al., 2003; Owen and Sherratt, 1998; Shvartsman et al., 2001).

**Figure 2**

## 2.1.    Network Model Balance Equations

We have modeled the molecular network as a *mass balance kinetic* model with variables describing the state of the system and rate constants. The evolution over time of a variable (molecular concentration) is represented by ordinary differential equations. Our network model consists of 13 state variables and 30 constants. The autocrine activation of the TGFα transcription activation and inhibition of TGFα-EGFR phosphorylation by PLCγ is modeled as an enzyme-catalyzed Michaelis-Menten reaction. The overall layout of the interaction network is described in **Figure 2**. Each molecular species is represented for simplicity as a variable $X_n$, where $n$ is its identifier, as listed in **Table 1**.

**Table 1**

Each of the reactions is briefly described here and is primarily based on peer-reviewed work published in the literature. Parameters and the source for the values are listed in **Table 2**.

**Table 2**

To summarize the reactions modeled, the EGFR-ligand-binding module includes the ligand TGFα ($X_1$) which binds to the receptor EGFR ($X_2$) and rapidly dimerizes to 2TGFα-EGFR ($X_3$), as modeled by Starbuck and Lauffenburger (1992). $X_3$ is then auto-phosphorylated to 2ppTGFα-EGFR ($X_4$) and the complex is internalized, referred to as TGFα-EGFR$_i$ ($X_5$) (Brightman and Fell, 2000; Schoeberl et al., 2002). The $X_3$ phosphorylation rate is enhanced by intracellular glucose ($X_{13}$) (Hertel et al., 1986; Steinbach et al., 2004), which was taken up from the environment. These processes are represented by **Eq. (1-4)**:

$$\frac{dX_1}{dt} = k_{-1} \cdot X_3 - k_1 \cdot X_1 \cdot X_2 + k_9 \cdot X_7, \tag{1}$$

$$\frac{dX_2}{dt} = k_{-1} \cdot X_3 - k_1 \cdot X_1 \cdot X_2 + k_8 \cdot X_5 - k_{-8} \cdot X_2, \tag{2}$$

$$\frac{dX_3}{dt} = 2 \cdot k_1 \cdot X_1 \cdot X_2 - 2 \cdot k_{-1} \cdot X_1 - k_2 \cdot X_3 \left[1 + w_g \cdot X_{13}\right] - k_3 \cdot X_3 + k_{-2} \cdot X_4 + \frac{V_{M2} \cdot X_{11}}{K_{M2} + X_{11}} \cdot X_4,$$

(3)

$$\frac{dX_4}{dt} = k_2 \cdot X_3 \left[1 + w_g \cdot X_{13}\right] - k_{-2} \cdot X_4 - k_4 \cdot X_4 - \frac{V_{M2} \cdot X_{11}}{K_{M2} + X_{11}} \cdot X_4.$$

(4)

The cytoplasmatic $X_5$ complex, in itself inactive (French and Lauffenburger, 1997), dissociates reversibly to cytoplasmic TGFα ($X_6$) and EGFR ($X_7$) as represented by **Eq. (5-7)**:

$$\frac{dX_5}{dt} = k_3 \cdot X_3 + k_4 \cdot X_4 + 2 \cdot k_{-5} \cdot X_6 \cdot X_7 - 2 \cdot k_5 \cdot X_5,$$

(5)

$$\frac{dX_6}{dt} = k_5 \cdot X_5 - k_{-5} \cdot X_6 \cdot X_7 - k_8 \cdot X_6 + k_{-8} \cdot X_2 + k_{12} \cdot X_8 - k_6 \cdot X_6,$$

(6)

$$\frac{dX_7}{dt} = k_5 \cdot X_5 - k_{-5} \cdot X_6 \cdot X_7 - k_{10} \cdot X_7 + k_{15} \cdot X_9 - k_7 \cdot X_7.$$

(7)

It has also been shown that increased internalization of $X_3$ and $X_4$ leads to down-regulation of EGFR RNA ($X_9$) expression and thus, diminished protein content (Hamburger et al., 1991). Conversely, EGFR activation by ligand binding increases TGFα RNA ($X_8$) synthesis. Both RNA species are also being transcribed and translated at a constitutive rate (Maruno et al., 1991; van der Valk et al., 1997) and both, protein and RNA are constantly degraded (Mader, 1988) as denoted by **Eq. (8, 9)**:

$$\frac{dX_8}{dt} = k_{13} \cdot C_1 - k_{14} \cdot X_8,$$

(8)

$$\frac{dX_9}{dt} = k_{17} \cdot C_1 - k_{16} \cdot X_9 + V_1 \cdot X_4,$$

(9)

where $C_1$ is the constant pool of nucleotides. The increased phosphorylated TGFα-EGFR surface complex raises the rate of transition from inactive PLCγ ($X_{10}$) to active PLCγ ($X_{11}$); this active PLCγ in turn exhibits negative feedback inhibition of $X_4$. (Chen et al., 1994; Chen et al., 1996; Wells, 1999) as represented by **Eq. (10, 11)**:

$$\frac{dX_{10}}{dt} = k_{21} \cdot X_{11} - k_{20} \cdot \left[C_2 - X_{11}\right] \cdot X_4, \tag{10}$$

$$\frac{dX_{11}}{dt} = k_{20} \cdot \left[C_2 - X_{11}\right] \cdot X_4 - k_{21} \cdot X_{11}, \tag{11}$$

where $C_2$ is the constant total PLCγ concentration. The intra-cellular glucose concentration ($X_{13}$) is increased by uptake (Noll et al., 2000) from the extracellular pool ($X_{12}$) and depleted by TGFα-EGFR phosphorylation (**Eq. (4)**) (Hertel et al., 1986; Steinbach et al., 2004), as described in **Eq. (12)**:

$$\frac{dX_{13}}{dt} = k_{23} \cdot X_{12} - k_2 \cdot X_3 \cdot X_{13}. \tag{12}$$

## 2.2.    Cellular Behavior

### 2.2.1.    Sub-cellular Molecular Flow

The concentration of a molecule in a given sub-cellular compartment ($X_m$) over time is expressed by **Eq. (13)**:

$$\frac{dX_j}{dt} = k_{in} \cdot \left[X_{j-1} + X_{j+1}\right] - 2 \cdot k_{out} \cdot X_j, \tag{13}$$

where $X_{j-1}$ is the concentration of the same molecule in the neighboring compartment before $X_j$ and $X_{j+1}$ is the compartment after $X_j$ where $j$ (1 to 4) is the compartment number and $k_{in}$ and $k_{out}$ are the flux rate constants into and out of the compartment, respectively (as shown in **Figure 1b**). At every time point at which the reaction network (**Eq. (1-12)**) is being calculated, their flux rates cause a redistribution based on mass-action, thus providing for the possibility of dynamic *spatial heterogeneity*.

### 2.2.2. Cell Migration

As stated earlier, using breast cancer cells Dittmar et al. (2002) could show that PLCγ is activated transiently and to a greater extent during migration and more gradually in the proliferative 'mode'. We adopt a simple threshold, $\sigma_{PLC}$, to decide whether the cell should undergo migration or not. Thus each cell is evaluated for its migratory potential (*M*), as given by **Eq. (14)**:

$$M_n\left[X_{11}\right] = \left[\frac{dX_{11}}{dt}\right]_n ,\qquad(14)$$

where $dX_{11}/dt$ is the change in concentration of PLCγ over time (*t*) and $n$ is the cell number. If $M_m > \sigma_{PLC}$ the *phenotypic decision threshold* is exceeded and the cell becomes eligible to migrate ($k_{25}$ in **Figure 2**). Evidence from previous studies on the EGFR autocrine signaling network points mainly to *local* factors being responsible for migration (Shvartsman et al., 2001) and as such, our virtual cells here evaluate the grid points within a *von Neumann neighborhood* for suitability. The mechanism that determines *where* the cell will migrate is decided by the spatial localization of the maximal active PLCγ within a cell. This is based on the findings from human breast adenocarcinoma cells in which EGF-induced cell migration is accompanied by the accumulation of PLCγ at the leading edge of migrating cells (Piccolo et al., 2002). Specifically, a migrating tumor cell then evaluates the intracellular concentration of PLCγ in the compartments that point towards adjacent locations to the North, East, South, and West of the cell's current site. The cell then selects one unoccupied lattice site among these four neighbors with a probability that depends on *both* the maximal concentration of active PLCγ at the leading edge, and the level of the so-called search precision, which we define below. This process is expressed in terms of a local valuation function for each neighboring lattice point *j* as given in **Eq. (15)**:

$$L_j\left[X_{11}, n_j\right] = \left(1 - n_j\right) \cdot \left[X_{11}\right]_j - \arg\max_{j \leq m}\left[X_{11}\right]_j \cdot \Psi_{PLC},$$ (15)

where $L_j$ is the value of a grid point in the neighborhood of the cell; the neighboring grid locations and compartments are numbered as *j* (1 to 4). In a given compartment *j*, $X_{11}$ is the concentration of activated PLCγ, *n* is the number of cells at that location in the neighborhood and *m* is the total number of compartments (here: *m*=4). $\Psi_{PLC} \in [0,1]$ represents the *search-precision* parameter that for a given run is held constant for all cells and corresponds to the *accuracy* of the cell's receptor-mediated direction-sensing mechanism as described in our previous work (Mansury and Deisboeck, 2003). Typically we set $\Psi_{PLC} = 0.7$, based on the same previous work in which we found that this value leads to the highest average velocity of the tumor's spatial expansion. If $L_j \geq 0$ and $n_j = 0$ (i.e., that location is unoccupied), then location *j* becomes eligible for the evaluating cell to migrate into it. If there are multiple locations that satisfy these two conditions, then the virtual cell *randomly* selects the next location. Note that the search precision parameter $\Psi_{PLC} = 0$ corresponds to a pure random walk, while $\Psi_{PLC} = 1$ means that cells never commit 'mistakes' and always migrate fully biased to the 'best' location with the highest level of PLCγ. To see this, consider $\Psi_{PLC} = 0$. In this case, the right-hand side of **Eq. (15)** is always non negative, which means all locations in the cell's neighborhood are eligible for migration. By contrast, when $\Psi_{PLC} = 1$, then cells always migrate to those locations exhibiting the highest level of PLCγ. As such, this search precision parameter determines how *sensitive* the migratory mechanism is to differences in active PLCγ concentration.

### 2.2.3. Cell Proliferation

If the change in concentration of active PLCγ is below the migration-threshold, $\sigma_{PLC}$, yet above a set noise threshold, $\sigma_n$ ($k_{26}$ in **Figure 2**), then the tumor cell will not chose to migrate yet has the potential to proliferate. However, an additional condition for proliferation is that the total cellular concentration of phosphorylated TGFα-EGFR exceeds a certain threshold $\sigma_{EGFR}$ ($k_{27}$ in **Figure 2**). Thus proliferation occurs if the proliferative potential $P_{prolif} \geq 0$. This potential is then calculated as given by **Eq. (16)**,

$$P_{prolif}\left[X_4\right] = X_4 - \sigma_{EGFR} \tag{16}$$

where $X_4$ is the concentration of ligand bound phosphorylated TGFα-EGFR complex in a cell. This function is derived from experimental observations citing cell proliferation in relation to an EGF-receptor threshold (Knauer et al., 1984) and a model, which relates receptor occupancy to percent maximal proliferation (Lauffenburger and Linderman, 1996). Specifically, these works report experimentally measured values of the cell proliferation response to EGFR occupancy for some human and rodent cell lines which demonstrated that $\sigma_{EGFR}$ is reached at 25% of the total receptor concentration.

It is noteworthy that we impose a limit on the number of cells that can proliferate in a given time period based on the Gompertz growth curve (Marusic et al., 1994). Furthermore, once a cell has been committed to proliferate, the newly divided cell will occupy one of the randomly chosen empty lattice sites in the *von Neumann* neighborhood. The time taken for division is delayed by ten iterations. This delay is motivated by the experimental finding that typical *cell cycle times* of glioma cells are approximately 26 hours, while the maximal migration rate is ~20 μm/hour (Hegedus et al., 2000). The scaled size of one of our lattice grid points is ~20 μm and so expansion due to cell proliferation over one grid point requires an order of magnitude (i.e., 26 times) more time than migration-driven expansion.

### 2.2.4. Cell Quiescence

For a cell to transition to a quiescent phenotype, it has to fulfill the following conditions: (*i*) a decline in PLCγ-concentration over time below the threshold $\sigma_{PLC}$, and (*ii*) a concentration of 2ppTGFα-EGFR of less than $\sigma_{EGFR}$. Under these conditions, the cell neither divides nor proliferates and we refer to this phenotype in our model as quiescent. *Cell death* or apoptosis is currently not included in our model.

### 2.3. Extracellular Grid

We employ a discrete lattice grid that represents a virtual slice of brain tissue. Specifically, the extracellular environment is modeled as a uniform 2D space consisting of a grid with 200 x 200 points in size. Each grid point can be occupied by only one cell at the same time. One single distant source of replenished nutrients, simulating the anatomical equivalent of a cross-sectional blood vessel, is located in the North-Eastern (NE) quadrant of the grid. This nutrient, represented by glucose, diffuses at a fixed rate uniformly over the lattice ($X_{12}$). Its flux ($J_s$) follows Fick's First Law of Diffusion as given by **Eq. (17)**,

$$J_s = -D \cdot \frac{\partial C_s}{\partial x} \cdot \frac{1}{n+1} \tag{17}$$

where $C_s$ is the concentration, $x$ distance in one dimension, $D$ is the diffusion coefficient of the molecule for the medium, and $n$ is the number of cells at a given location. Since we allow only one cell per lattice point, if a cell is present at that point ($n=1$), the flux due to diffusion is reduced by half.

We assume *Dirichlet boundary* conditions for the diffusing glucose, where the value is fixed at zero at the edges. The autocrine secreted protein growth factor TGFα is also deposited on grid points in the neighborhood of a cell at the rate given by **Eq. (1)** and **Table 2.** It is only replenished if a cell is located in an adjacent grid point. Thus TGFα is an *autocrine* produced hormone which can, in addition, act in a *paracrine* as well as *juxtacrine* manner. Previous work has shown that the capture time for autocrine ligands of EGFR is extremely short (Shvartsman et al., 2001), and as such we can treat it for now as not diffusing, rather as a residual chemical 'track' marking a lattice location previously occupied by a cell. Finally, this TGFα protein outside the cell is also assumed to have a rate of degradation.

# 3. Results

Our code is implemented in Java (Sun Microsystems, Inc., USA) and uses the RePast (version 2.0) agent-based modeling toolkit (http://repast.sourceforge.net), combined with in-house developed classes for representing molecules, reactions and sub-cellular compartments as a set of hierarchical objects. For a typical simulation run with three different random number seeds and scanning seven parameter values (21 runs) the algorithm required 18 hrs 46 min of CPU time on a computer with dual Intel Xeon 2.3GHz processors, connected via gigabit Ethernet to the central file storage system and running Linux.

## 3.1. Multi-cellular Dynamics

We simulate the expansion of the multicellular tumor from its initial central seed towards the peak of glucose located in the NE quadrant of the grid. The time it takes for the first migrating cell to reach the edges of this peak is used as a measure of the tumor system's *spatio-temporal expansion dynamics*. **Figure 3a** demonstrates that when the PLCγ-dependent cell decision threshold $\sigma_{PLC}$ is very low, the spatio-temporal expansion of the tumor is accelerated as the cancerous system requires less time to reach the source. Increasing this threshold (decreases the probability of a cell attaining the migratory phenotype and thus) slows the tumor system down until, beyond a minimum expansion velocity at a $\sigma_{PLC}$ of approximately $3.5 \times 10^{-3}$ nM/s, the system plateaus at a $\sigma_{PLC}$ of roughly $5 \times 10^{-3}$ nM/s. Correspondingly, as $\sigma_{PLC}$ increases, the ratio of migrating to proliferating cells decreases, in fact approaching zero when no cell migration occurs anymore beyond a $\sigma_{PLC}$ of $\sim 5 \times 10^{-3}$ nM/s (**Figure 3b**). Combined, these two figures confirm that a smaller proportion of migrating cells yields slower rates of *overall* tumor expansion.

**Figure 3 a**

**Figure 3 b**

## 3.2. Sub-cellular Dynamics

The 2D snapshots at four consecutive time points depict the spatial patterns of the three cellular phenotypes, i.e., proliferative, migratory and quiescent tumor cells. The migratory decision threshold ($\sigma_{PLC}$) was set at 0.001 to ensure stable phenotypes and fast expansion (**Figure 4a**) with a ratio of migratory to proliferative cells approximating five (compare with **Figure 3b**) at the endpoint of the run which is reached when the first migrating cell enters the edge of the NE quadrant (**Figure 4b**). The plots describe a *mixed-phase* of expansion, where *both* phenotypic traits, i.e. migration *and* proliferation occur within the cancerous cell population.

**Figure 4 a**

**Figure 4 b**

**Figure 4 c**

On the *molecular* level, we first note that $X_1$ (TGFα protein) is *homogenously* deposited in the extracellular von Neumann neighborhood (**Figure 4c**). The sub-cellular *profile* of protein components of the EGFR-network within this 'first' migratory cell shows that $X_4$ (2pp-TGFα-EGFR) is also *homogeneously* distributed in the cell membrane in all directions. Conversely, $X_{11}$, i.e., the concentration of active PLCγ displays a *polarized* pattern within the cytoplasm. Specifically, the maximal PLCγ concentration [i.e., 0.16 nM] resides in the cytosolic compartment *closest* to the NE quadrant where the glucose source is located.

## 4. Discussion and Conclusions

Experimental evidence suggests that a molecular switch operates between the phenotypes of proliferation and migration in highly malignant brain tumor cells. To investigate this behavior further, we have integrated here a sub-cellular decision-making gene-protein network into a previously developed multiscale, agent-based modeling environment. The results demonstrate that this combined *molecular-microscopic-macroscopic* algorithm is capable of producing ranges of behavior at distinct scales, solely by varying the value of the molecular parameter, $\sigma_{PLC}$. Specifically, lowering this *cellular* phenotypic decision threshold of phospholipaseC-$\gamma$ leads to fast *multicellular* tumor expansion while raising the threshold value yields slower spatial expansion rates, conferred by a smaller portion of migratory cells within the system. Interestingly, this behavior is not smooth; rather it indicates a *phase transition* at a $\sigma_{PLC}$ of roughly 2.5 x $10^{-3}$ nM/s. One can argue that this is an *emergent* property of the system since no *a priori* condition in the algorithm forces such behavior. Similarly, at the *single cell* level, the *polarized* localization of active PLC$\gamma$ in the first migratory cell to reach the edge of the glucose source is also an emergent phenomenon, since the flux constants for each compartment are identical. The latter is consistent with previous experimental work, in which EGFR-activation-dependent migration of human adenocarcinoma cells was accompanied by translocation of PLC$\gamma$ to the leading edge (Piccolo et al., 2002). In addition, our model predicts a lack of polarization in EGFR ligand-receptor complex at the membrane at least for this 'first' migratory cell. While this appears to be in accordance with previous reports (Bailly et al., 2000) it requires further, more detailed investigation of different phenotypes as well as of cells at other locations and at different time points in the system.

The ability of our system to simulate tumor growth over several orders of magnitude driven by a decision making gene-protein network allows us to examine both the *molecular signature* that defines the phenotypic switch as well as the multicellular patterns that in turn may serve as a *macroscopic systems 'read-out'* for dynamical changes in its molecular profiles. At the moment, our virtual biosystem operates with only two genes, i.e. TGF$\alpha$ and EGFR, due to the data-rich nature of this network. However, the extensibility of the platform will allow us to add multiple genetic interaction networks and simulate their behavior over multiple scales. A comparison of such simulations with experimental data generated using e.g. cDNA and oligonucleotide arrays to study gliomas (e.g., Mariani et al., 2001; Sallinen et al., 2000) is likely to yield a powerful virtual discovery platform through (*i*) designing *in silico* experiments, (*ii*) developing novel hypotheses, and (*iii*) testing them by guiding specific experimental work which can provide further modeling input. It is thus essential that most of the predicted dynamics of molecular species concentration and localization in our modeling platform, at both the cellular and multicellular level, are *experimentally testable*. For example, the simulation output of the spatial localization of the proteins can be queried at every single time point by immuno-staining or even live cell fluorescence microscopy and proteomic approaches to test their phosphorylation status. Compartmental flux rates for cell surface and cytoplasmic molecules can also be experimentally estimated using methods like fluorescence recovery after photobleaching (FRAP) and fluorescence loss in photobleaching (FLIP) in living cells (Axelrod et al., 1976). Lastly, receptor association and dissociation rates have already been measured mostly by radioactively labeled ligand binding, followed by model fitting to estimate the parameters of interest (French et al., 1995).

However, the signaling model adopted in this first approximation here is admittedly simple by design. For instance, we do not yet include explicitly any components of the ERK pathway, nor do we consider the signaling downstream of PLC$\gamma$ or the Ca$^{2+}$ signaling that is associated with it. In addition, at the cellular level, there is currently no treatment of mechanical or pH-related aspects to tumor growth either. In future work we will therefore step-wise add other relevant aspects including environmental factors such as mechanical stress, known to restrict the expansion of tumors *in vitro*, in an effort to study the interaction of the chemical and physical interactions and their effect on the tumor growth dynamics at various scales.

Nonetheless, already in its present form, we argue that this platform is an important first step in realizing a fully validated multiscale molecular *and* multicellular *in silico* model of tumor growth. If combined with proper experimental input, this algorithm will prove useful in improving our understanding of tumor biology, not limited to brain tumors, and has the potential to help guide future research in the quest for designing and developing more effective molecular anti-cancer therapeutics, *with* systems impact.

## Acknowledgements

This work has been supported in part by NIH grant CA 085139 and by the Harvard-MIT (HST) Athinoula A. Martinos Center for Biomedical Imaging and the Department of Radiology at Massachusetts General Hospital. Y.M. is the recipient of an NCI-Training Grant Fellowship from the National Institutes of Health (CA 09502).

## Table and Figure Legends

**Table 1.** Shown here are the variables of the network model and the molecular species they represent. The table includes the initial values in [$nM$] and the inter-compartmental flux rates [$s^{-1}$] with their respective literature sources. Reasonable estimates were used where no published values were available.

**Table 2.** Listed are the symbols used for the parameters of the equations of the network described both in text and **Figure 2** as well as their values and the related literature sources. All first order rate constants are listed in terms of [$s^{-1}$], second order in [$M^{-1}s^{-1}$] and cooperative reaction constants are given in terms of [$nM$].

**Figure 1.** The spatial compartmentalization of a tumor cell is depicted schematically. **(a)** Each cell consists of a central nucleus, surrounding cytoplasm and a membrane compartment. These compartments are then divided into four sub-compartments in the cardinal directions, each connected to two others. The *gray* sub-compartmental region is further detailed in **(b)**. Here, the intra-compartmental flux consists of a rate of inflow (**$k_{in}$**) and outflow (**$k_{out}$**) as represented by solid arrows, while the stippled arrows indicate the exchange of mass between different compartments as a result of the gene-protein interaction network (see **Figure 2**).

**Figure 2.** The figure depicts the implemented gene-protein interaction network of the TGFα-EGFR signaling pathway. Each arrow represents a reaction that is in turn represented by a rate constant referred to in **Table 2**. The arrows that start from a molecule species and terminate in the environment signify rates of degradation whereas those with stippled lines with either plus (+) or minus (-) signs indicate positive and negative feedback regulation,

respectively. The *gray* arrows represent the cell's phenotypic 'decision' of entering into a proliferative or migratory 'mode'.

**Figure 3.** The plots describe the effect of varying $\sigma_{PLC}$ (*x-axis*) in [*nM/s*] on (**a**) the time in [*min*] it takes for the first migratory tumor cell to reach the source of glucose in the NE quadrant (*y-axis*), and on (**b**) the ratio of migrating to proliferating cells (*y-axis*) within the tumor system. The error bars indicate the standard deviation between ten runs using different random number seeds.

**Figure 4.** The figure shows the spatial expansion of the tumor in a single run with $\sigma_{PLC} = 0.001$. (**a**) The 2D tumor snapshots display migrating (*black*), proliferating (*white*), and quiescent cells (*gray-striped)* as well as the empty lattice grid (*light-gray*) and grid sites with the diffusing nutrient glucose (*dark-gray*). The (*red*) circle in the NE quadrant indicates the initial location of the nutrient source from which glucose diffuses. Depicted is the tumor progression at time points $t = 50, 300, 400$ and $502$; the migrating tumor cell that first enters the edge of the glucose source is highlighted (*yellow*) in (**b**) and further magnified. (**c**) The sub- and extracellular localization of three molecular protein species within this 'first' migratory cell is plotted in 2D. These proteins include extracellular TGFα (*dark-red*) in the von Neumann neighborhood, phosphorylated TGFα-EGFR located in the cell membrane and active PLCγ within the cytosol. The color-bar indicates the concentration range in [*nM*] of the molecular species with *dark-blue* depicting zero (for cellular 'geography' compare with **Figure 1**). The *gray* arrow points to the location of the glucose source relative to the cell.

# Tables

## Table 1.

| Symbol | Variable | Initial Values [nM] | Sub-cellular Flux Rates | Reference |
|--------|----------|---------------------|-------------------------|-----------|
| $X_1$ | TGFα extracellular protein | 1 | $5 \times 10^{-2}$ | (Dowd et al., 1999) |
| $X_2$ | EGFR cell surface receptor | 25 | $1 \times 10^{-4}$ | (Maly et al., 2004) |
| $X_3$ | Dimeric TGFα-EGFR cell surface complex | 0 | $1 \times 10^{-4}$ | (Maly et al., 2004) |
| $X_4$ | phosphorylated active dimeric TGFα-EGFR cell surface complex | 0 | $1 \times 10^{-4}$ | (Maly et al., 2004) |
| $X_5$ | Cytoplasmic inactive dimeric TGFα-EGFR complex | 0 | $1 \times 10^{-2}$ | (Hirschberg et al., 1998) |
| $X_6$ | Cytoplasmic EGFR protein | 0 | $1 \times 10^{-2}$ | (Hirschberg et al., 1998) |
| $X_7$ | Cytoplasmic TGFα protein | 1 | $1.5 \times 10^{-2}$ | (Hirschberg et al., 1998) |
| $X_8$ | EGFR RNA | 1 | $1 \times 10^{-2}$ | (Kues et al., 2001) |
| $X_9$ | TGFα RNA | 0 | $2 \times 10^{-2}$ | (Kues et al., 2001) |
| $X_{10}$ | PLCγ Ca-bound | 1 | No-flux modeled | (Piccolo et al., 2002) |
| $X_{11}$ | PLCγ active, phosphorylated, Ca-bound | 1 | $2 \times 10^{-4}$ | (Kim et al., 1990) |
| $X_{12}$ | Nucleotide pool | 5 | No-flux modeled | Estimate |
| $X_{13}$ | Glucose cytoplasmic | 1 | $3 \times 10^{-3}$ | (Pfeuffer et al., 2000) |
| $X_{14}$ | Glucose extracellular | 0 | $3 \times 10^{-5}\ min^{-1}$ | (Jain, 1987) |

## Table 2.

| Parameter | Value | Description | Reference |
|-----------|-------|-------------|-----------|
| $k_1$ | $3 \times 10^7$ | TGFα-EGFR cell-surface complex formation rate | (De Crescenzo et al., 2000; Kramer et al., 1994; Rutten et al., 1996) |
| $k_{-1}$ | $3.8 \times 10^{-3}$ | Rate of dissociation of TGFα-EGFR cell-surface complex | (De Crescenzo et al., 2000; Kramer et al., 1994; Rutten et al., 1996) |
| $k_2$ | $1 \times 10^{-3}$ | Rate of TGFα-EGFR phosphorylation | (Brightman and Fell, 2000) |
| $k_{-2}$ | $1 \times 10^{-6}$ | Rate of TGFα-EGFR de-phosphorylation | (Brightman and Fell, 2000) |
| $k_3$ | $5 \times 10^{-5}$ | Rate of phosporylated TGFα-EGFR internalization | (Schoeberl et al., 2002; Starbuck and Lauffenburger, 1992) |
| $k_4$ | $5 \times 10^{-5}$ | Rate of cell-surface TGFα- | (Starbuck and Lauffenburger, 1992) |

| | | EGFR internalization | |
|---|---|---|---|
| $k_5$ | $1 \times 10^{-2}$ | Dissociation rate of cytoplasmic TGFα-EGFR | (Schoeberl et al., 2002; Starbuck and Lauffenburger, 1992) |
| $k_{-5}$ | $1.4 \times 10^5$ | Reverse dissociation rate of cytoplasmic TGFα-EGFR | (Schoeberl et al., 2002; Starbuck and Lauffenburger, 1992) |
| $k_6$ | $1.67 \times 10^{-4}$ | Rate of cytoplasmic EGFR protein degradation | (French and Lauffenburger, 1997; Wiley and Cunningham, 1981) |
| $k_7$ | $1.67 \times 10^{-4}$ | Rate of cytoplasmic TGFα protein degradation | (French and Lauffenburger, 1997; Wiley and Cunningham, 1981) |
| $k_8$ | $5 \times 10^{-3}$ | Rate of cytoplasmic EGFR insertion into the membrane | (Schoeberl et al., 2002; Starbuck and Lauffenburger, 1992) |
| $k_{-8}$ | $5 \times 10^{-5}$ | Rate of cell-surface EGFR internalization | (Starbuck and Lauffenburger, 1992) |
| $k_9$ | 1 | Rate of membrane insertion and secretion of TGFα | (Borrell-Pages et al., 2003; Shvartsman et al., 2002a; Tang et al., 1997) |
| $k_{10}$ | 0.01 | Rate of down-regulation of EGFR expression by the TGFαEGFR complex | (Hamburger et al., 1991) |
| $k_{11}$ | 0.01 | Degradation of extracellular TGFα | Estimate |
| $k_{12}$ | 5 | [molecules/min] Rate of translation of EGFR RNA | (Mader, 1988) |
| $k_{13}$ | 2.17 | Basal transcription rate EGFR RNA [molecules/min] | (Mader, 1988) |
| $k_{14}$ | $1.2 \times 10^{-3}$ | EGFR RNA degradation rate [molecules/min] | (Mader, 1988) |
| $k_{15}$ | 5 | Rate of translation of TGFα [molecules/min] | (Mader, 1988) |
| $k_{16}$ | $1.2 \times 10^{-3}$ | TGFα RNA degradation rate [molecules/min] | (Mader, 1988) |
| $k_{17}$ | 12 | Basal transcription rate TGFa_rna [molecules/min] | (Mader, 1988) |
| $k_{18}$ | $K_{M1}$, $V_{M1}$, $w_1$ | Induction of TGFα transcription by activated TGFα-EGFR at the cell surface | - |
| $K_{M1}$ | 1 | Km of TGFα RNA transcriptional activation | Estimate |
| $V_{M1}$ | 5 | Rate of TGFα RNA transcriptional activation | Estimate |
| $w_1$ | 1 | Weight of Hills Coefficient of TGFα RNA activation | Estimate |
| $k_{19}$ | 0.1 | Enhanced rate of PLCγ activation by EGFR | (Haugh et al., 2000) |
| $k_{20}$ | 0.1 | Basal rate of activation of PLCγ | (Haugh et al., 2000) |
| $k_{21}$ | 0.05 | Rate of in-activation of PLCγ | (Haugh et al., 2000) |
| $k_{22}$ | $K_{M2}$, $V_{M2}$, $w_2$ | PLCγ dependent rate of de-phosphorylation of phosphorylated TGFα-EGFR | - |
| $K_{M2}$ | 5 | Km PLCγ inhibition of phosphorylated surface TGFα-EGFR | Estimate |
| $V_{M2}$ | 0.25 | PLCγ inhibition rate of LR* | Estimate |
| $\mathbf{w_2}$ | 1 | Weight of hills coefficient PLCγ inhibition of phosphorylated surface TGFα-EGFR | Estimate |
| $\mathbf{k_{23}}$ | 0.7 | Lumped rate of Glucose uptake | (Noll et al., 2000) |
| $\mathbf{k_{24}}$ | 0.01 | Increased rate of TGFα-EGFR phosphorylation by Glucose | (Hertel et al., 1986; Steinbach et al., 2004) |
| $\mathbf{w_g}$ | 5.0 | Weight of increase in rate of TGFα-EGFR phosphorylation by Glucose | Estimate |
| $\mathbf{k_{25}}$ | Eqs. 14, 15 | Migratory signal | (Dittmar et al., 2002; El-Obeid et al., 1997; Kruger and Reddy, 2003) |
| $\mathbf{k_{26}}$ | Eq. 14 | Mitotic signal I | (Knauer et al., 1984; Maruno et al., 1991; Schoeberl et al., 2002) |
| $\mathbf{k_{27}}$ | Eq. 16 | Mitotic signal II | (Chen et al., 1994; Dittmar et al., 2002; Knauer et al., 1984; Kruger and Reddy, 2003; Maruno et al., 1991) |

**Figure 1.**

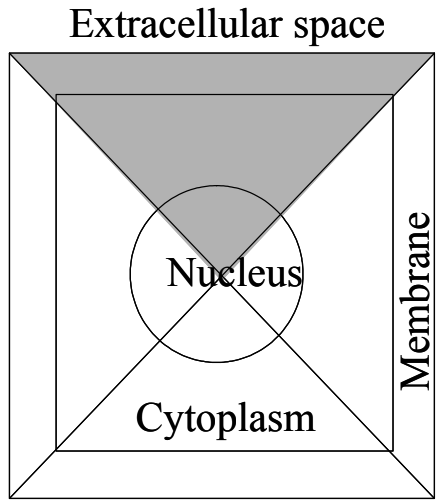

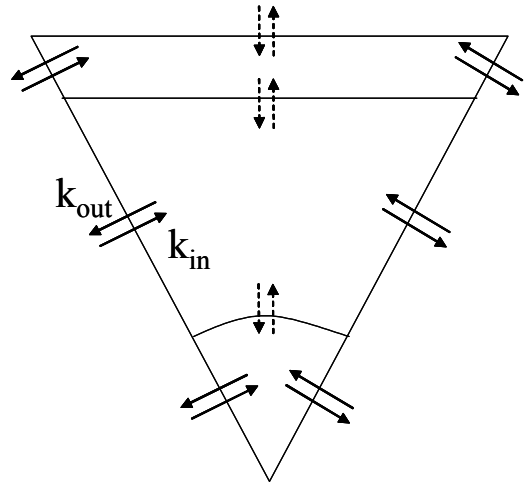

**(a)**                              **(b)**

**Figure 2.**

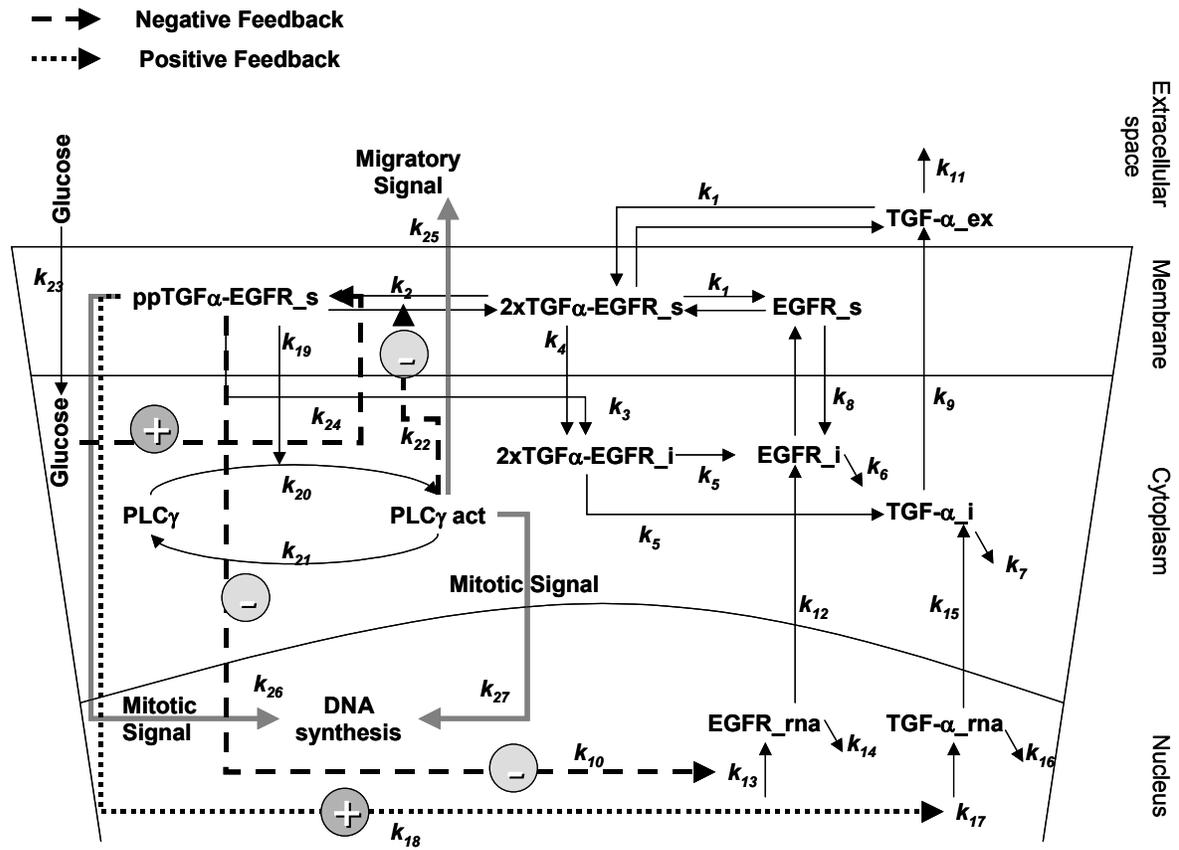

**Figure 3 (a).**

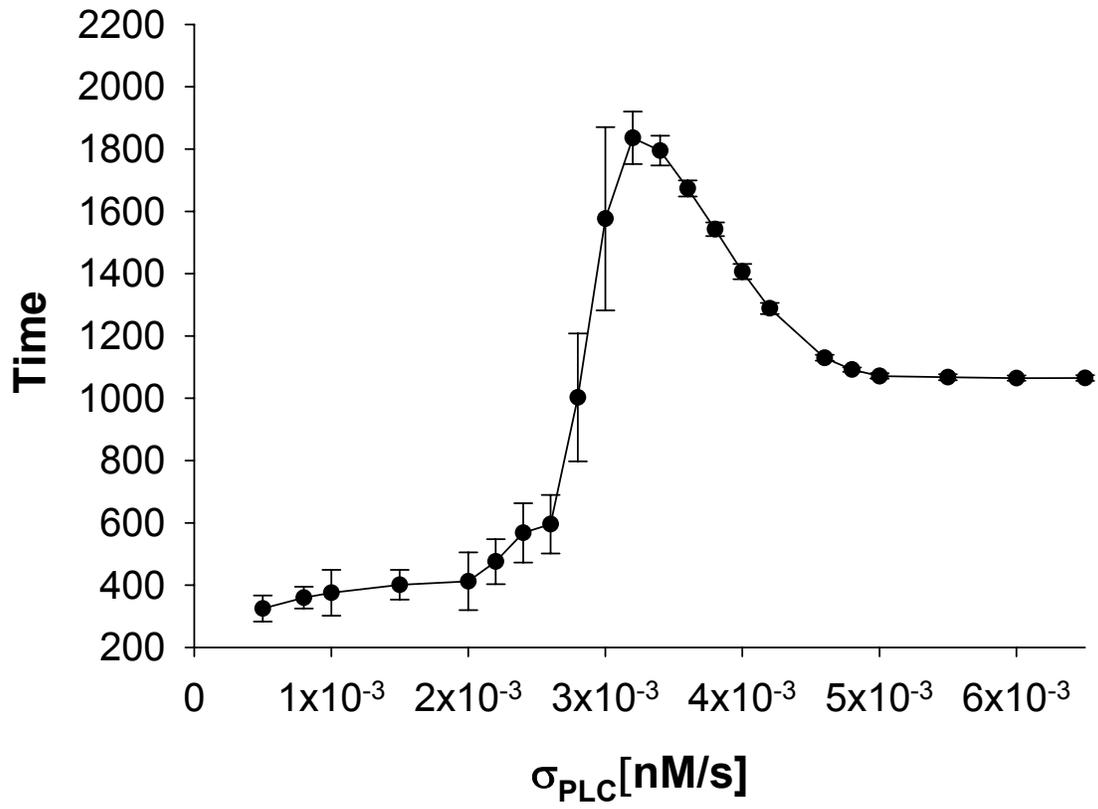

**Figure 3 (b).**

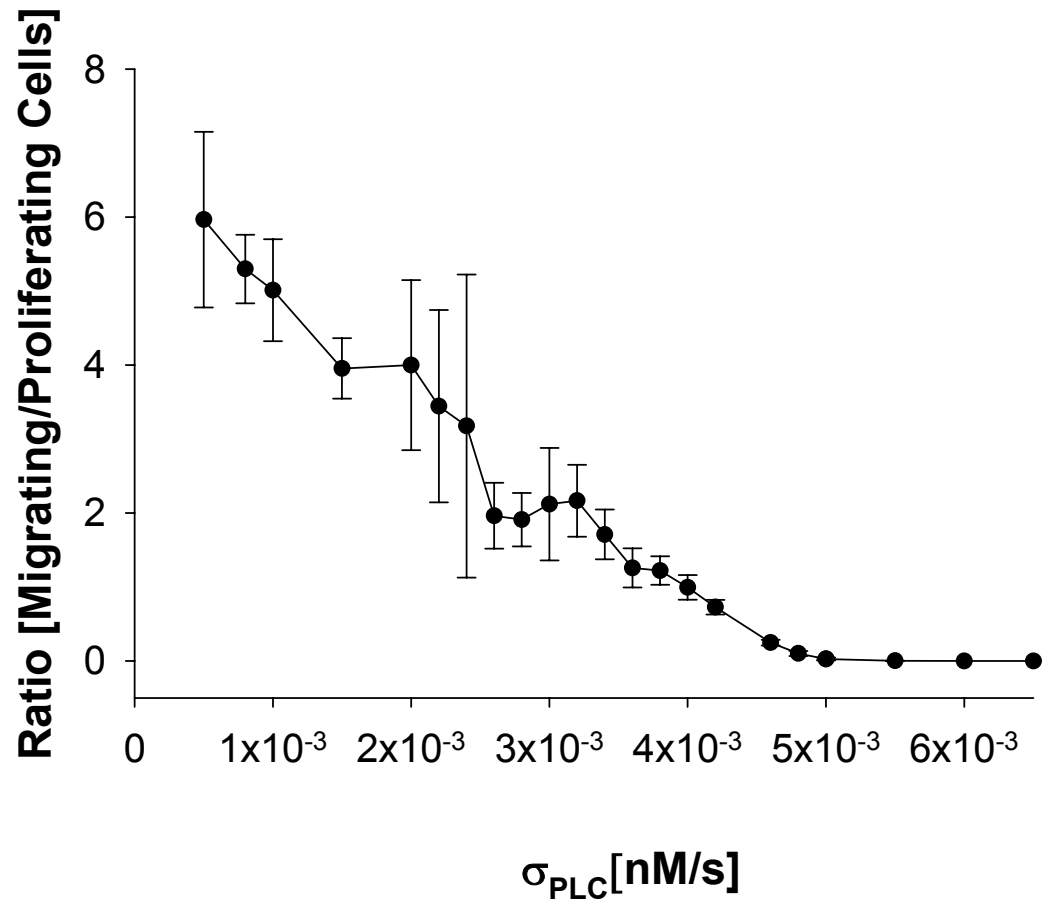

**Figure 4.**

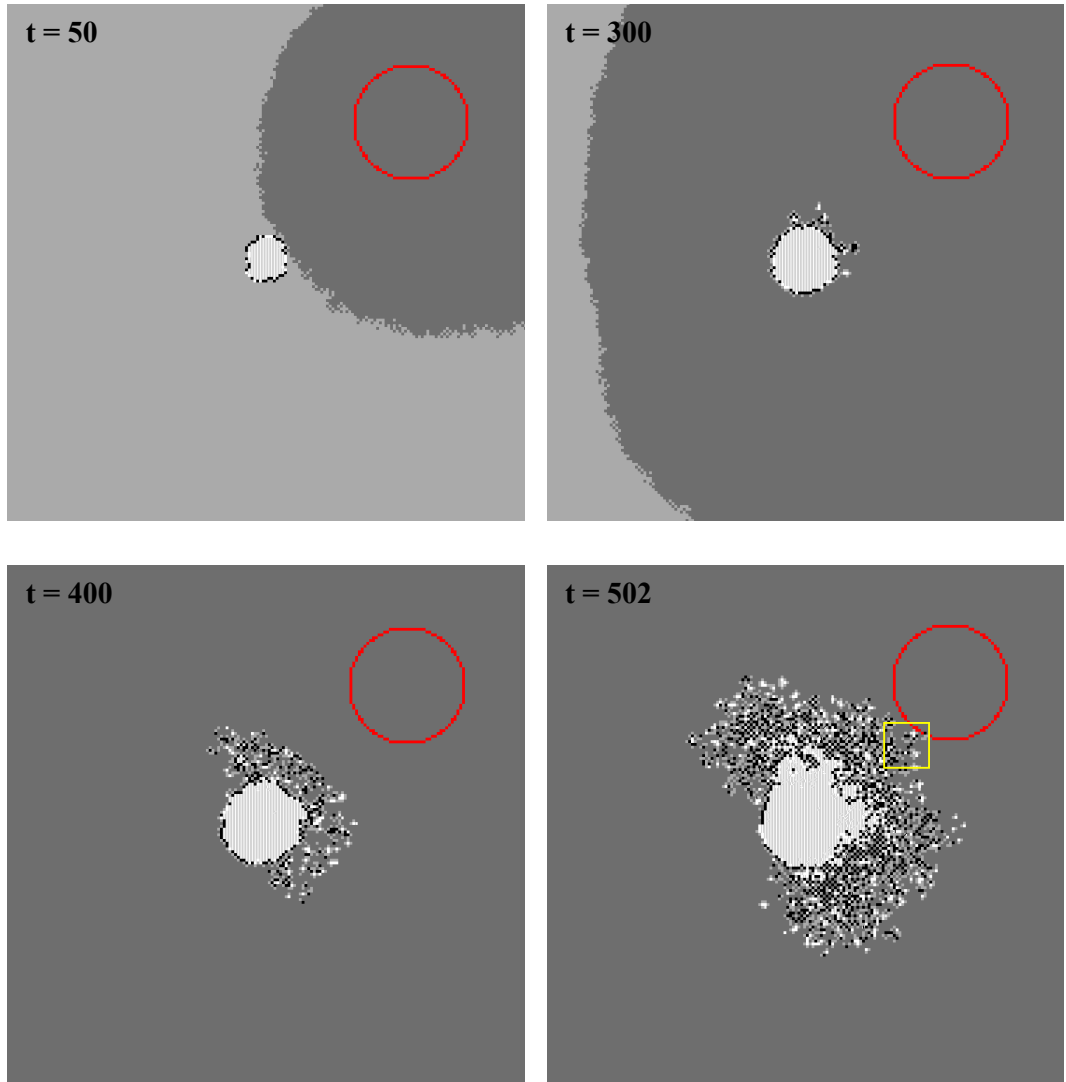

(a)

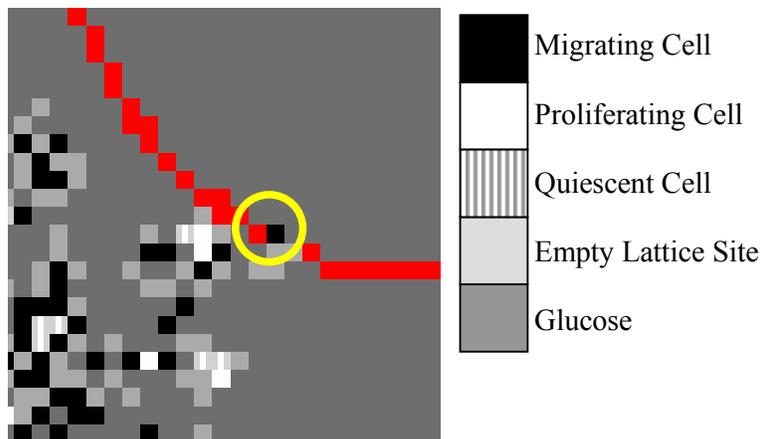

(b)

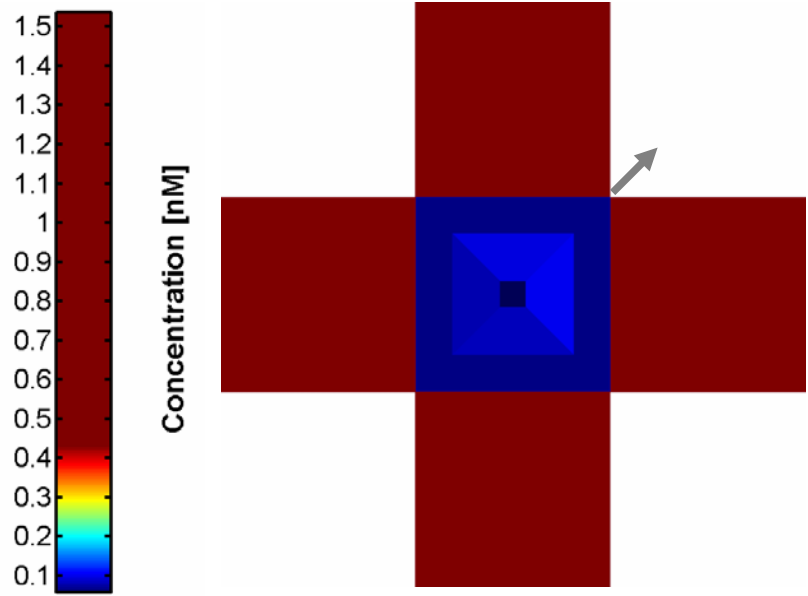

**(c)**

**Footnotes**

1.  In those instances where no glioma data were available in the literature, we have used published EGFR data derived from other relevant experimental systems, primarily from other human carcinoma cell lines.